\documentclass[a4paper,11pt]{article}
\usepackage{amsmath}
\usepackage{amssymb}
\usepackage{graphicx}
\begin{document}
\topmargin 0pt \oddsidemargin 0mm

\begin{titlepage}

\vspace{5mm}
\begin{center}
{\Large \bf A Self-gravitating Dirac-Born-Infeld Global Monopole } \vspace{20mm}

{\large Dao-Jun Liu\footnote{e-mail address:
djliu@shnu.edu.cn}},  {\large Ying-Li Zhang}  and
 {\large Xin-Zhou Li}

\vspace{1mm} {\em  Center for Astrophysics,
Shanghai Normal University, 100 Guilin Road, Shanghai
200234,China}

\end{center}

\begin{abstract}
We generalize the field theory of global monopole to Dirac-Born-Infeld(DBI) field and investigate the gravitational property of a DBI global monopole  in four-dimensional spherically symmetric spacetime. The coupled equations for the
metric and the DBI scalar field are solved asymptotically and numerically. It is  found that, just as a canonical global monopole, the gravitational effect of DBI global monopole is equivalent to that of a deficit solid angle in the metric
plus a negative mass at the origin. However, compared with a canonical global monopole, for the same false vacuum and symmetry breaking scale, a DBI global monopole has a relatively smaller core  and a larger absolute value of effective mass. Thus, it can give out a larger deflect angle when the light passing by. Especially, when the scale of warp factor is small enough, the effective mass of a DBI global monopole does not depend apparently on the value of the false vacuum, which is qualitatively different from that of a canonical global monopole.
\end{abstract}

\vspace{5mm}
\end{titlepage}

\newpage

\section{Introduction}
Different kinds of topological defects could be formed as a result
of symmetry-breaking phase transitions\cite{Kibble,Vilekin,Vilenkin-SHellard}.
In cosmology, these defects attract much attention since they might
appear naturally during phase transitions in early universe.
The symmetry breaking model of canonical scalar field \cite{Li}  is usually prototypically presented by
\begin{equation}
\mathcal{L}=\frac{1}{2}\partial_{\mu}\phi^{a}\partial^{\mu}\phi^{a}-V(f),
\end{equation}
where $\phi^a$ is a set of scalar fields, $a=1,\cdots,N$ and $f=(\phi^a\phi^a)^{1/2}$. The model has $O(N)$ symmetry and admits the domain wall, string and monopole solutions for $N=1,2$ and $3$, respectively. Usually, the potential $V(f)$ has the minimum at a finite non-zero value of $f$.

 Global monopole, which
arise in some models where a global $SO(3)$ symmetry is spontaneously
broken to $U(1)$, is one of the most interesting defects.
  Many important properties of global monopole with a canonical kinetic term (canonical global monopole)
have been well studied (see, for a review, \cite{Vilenkin-SHellard}). The most striking property of global monopole is that its mass is divergent in flat spacetime.  The gravitational effects of global
monopoles were firstly studied by Barriola and
Vilenkin\cite{Barriola}. When gravity is considered, the linearly divergent mass of the global monopole has an effect analogous to that of a deficit solid angle plus that of a tiny mass at the origin. Later, Harari and
Loust\'o\cite{Harari}, and Shi and Li\cite{Shi} realized that this small gravitational mass is actually negative. It is shown that the properties of global monopoles in asymptotically de Sitter/anti-de Sitter(dS/AdS) spacetime \cite{Hao} and Branes-Dicke theory \cite{Lu} are very different from those of an ordinary global monopole.

In recent years, field theories with non-canonical
fields were intensively investigated  in the cosmological context. The so-called $k$-fields were first introduced in the context of inflation (dubbed $k$-inflation scenario \cite{Armendariz-Picon}) and then $k$-essence models were suggested as a candidate to dark energy \cite{kessence}.
 Tachyon matter \cite{Sen} and ghost condensation \cite{ghost}
are other examples of non-canonical fields in cosmology.
Symmetry-breaking models of $k$-field theory have also been discussed in the literature \cite{liu}.
General properties of global topological defects appearing in such
models were studied in Ref.\cite{Babichev}.
It has been shown that the properties of such defects (dubbed k-defects) are quite different from "standard" global domain walls, vortices and monopoles. The gravitational property of a kind of global $k$-monopole was investigated in Ref.\cite{Jin}. It was shown that a gravitational global $k$-monopole can produce a tiny gravitational field outside the core in addition to a solid angular deficit and, as a new feature compared with the canonical global monopole, this gravitational field can be attractive or repulsive depending on the non-canonical kinetic term in the Lagrangian.

As a kind of $k$-field,  Dirac-Born-Infeld (DBI) field recently attracted much attention. The main reason is that, in some string theories, the motion of the brane seems to admit an effective
good description in terms of a DBI action
coupled to gravity \cite{DBI}, and it results in a scalar field theory
with non-canonical kinetic terms.
Such a DBI action with higher derivative terms gives a
variety of novel cosmological consequences \cite{DBI1}. The cosmological evolution of the system of a DBI field plus a perfect fluid was investigated \cite{Guo}.  Sarangi \cite{Sarangi} recently constructed global vortex solutions in the world volume of a $D3$-brane described by the DBI action.

In this work, we would like to generalize the field theory of global monopole to DBI field and investigate the gravitational property of a DBI global monopole  in four-dimensional spherically symmetric spacetime. It is found that, compared with a canonical global monopole, for the same false vacuum and symmetry breaking scale, a DBI global monopole has a smaller core  and the absolute value of its effective mass is larger, then it can give a larger deflect angle to the light passing by. Especially, when the scale of warp factor is small enough, the effective mass of a DBI global monopole does not depend apparently on the value of the false vacuum, which is qualitatively different from that of a canonical global monopole.

\section{The model}
The DBI action for the $D3$-brane on a warped background is given by \cite{DBI}
\begin{equation}
S=\int d^4{x}\sqrt{-g}\left[f(\phi)^{-1}\left(1-\sqrt{1-f(\phi)g^{\mu\nu}\partial_{\mu}\phi\partial_{\nu}\phi}\right)-V(\phi)\right],
\end{equation}
where $f(\phi)$ is the warp factor and $g_{\mu\nu}$ is the metric with signature $[+ - - -]$. To study monopole solutions, we must generalize the above action to the case of a $3$-fold scalar field.

To be specific, let us consider a particular model of the DBI $3$-fold scalar field which couples minimally to the Einstein's gravity, where a
global $O(3)$ symmetry is broken down to $U(1)$.
The system is described by the action (we work in units such that $c=\hbar=1$)
 \begin{equation}\label{S0}
S=\int
d^4{x}\sqrt{-g}\left(\frac{\mathcal{R}}{16\pi G}+\mathcal{L}_{DBI}\right),
\end{equation}
where $G$ is Newton's
constant, $\mathcal{R}$ denotes Ricci scalar  and the Lagrangian density of the DBI scalar field reads%
\begin{equation}\label{Lagrangian}
\mathcal{L}_{DBI}=
M^4\left(1-\sqrt{1-\frac{g^{\mu\nu}\partial_{\mu}\phi^a\partial_{\nu}\phi^a}{M^4}}\right)
-V_0\left(\phi^{a}\phi^a-\sigma_0^2\right)^2,
\end{equation}
where ${\phi}^a$ is the $SO(3)$ triplet of the DBI scalar field with the internal $O(3)$ index $a=1,2,3$, the scale of warp factor $M$ and the symmetry breaking scale $\sigma_0$ are two constants with a dimension of mass and $V_0$ is a dimensionless constant. In Eq.(\ref{Lagrangian}), $\phi^{1},\phi^{2},\phi^{3}$ could, for example, correspond to three directions in a five-dimensional warped throat geometry and the potential then corresponds to the potential for these direction locations of the $D3$-brane.

The hedgedog configuration describing a global monopole is given by
\begin{equation}\label{mono solution}
\phi^{a}=\sigma_0f(\rho)\frac{ x^a}{\rho},
\end{equation}
where $x^a x^a=\rho^2$. Therefore, we shall actually
have a global monopole solution if $f\rightarrow 1$ at spatial
infinity and $f\rightarrow0$ near the origin.

For an isolated static monopole, the metric must have spherical symmetry, so it is convenient to  work under the coordinates which make the metric reads
\begin{equation}\label{line element}
{d s^2}=B(\rho)dt^2-A(\rho)d\rho^2-\rho^2 (d\theta^2+\sin^2\theta
d\varphi^2),
\end{equation}
where the usual relation between the spherical coordinates $\rho$,
$\theta$, $\varphi$ and Cartesian coordinate $x^a$ is kept. The
equation of motion for $f(r)$ can be obtained from
Eqs.(\ref{Lagrangian})-(\ref{line element}) by introducing a new
dimensionless parameter  $r = \sigma_0 \rho$ :
\begin{eqnarray}\label{E-L Eq}
\frac{1}{A}\left(\gamma f^\prime\right)^\prime +\frac{\gamma
f^\prime}{A}\left[\frac{2}{r}+\frac{1}{2}\left(\frac{B^\prime}{B}-\frac{A^\prime}{A}\right)\right]= \frac{2\gamma
f}{r^2} + 4V_0 f\left(f^2-1\right),
\end{eqnarray}
where Lorentz factor $\gamma$ reads
\begin{equation}
 \gamma =
\frac{1}{\sqrt{1+\lambda\left(\frac{f^{\prime2}}{A}+\frac{2f^2}{r^2}\right)}} ,
\end{equation}
the dimensionless constant $\lambda$ is defined as
\begin{equation}
\lambda\equiv \frac{\sigma_0^4}{M^4}
\end{equation}
and the primes denote the derivatives with respect to $r$. Note that if the scale of warp factor $M\rightarrow \infty$, then Lorentz factor $\gamma\rightarrow 1$ and Eq.(\ref{E-L Eq}) reduces to the equation of motion for a canonical global monopole.

The Einstein equations  $G_{\mu\nu}=8\pi G T_{\mu\nu}$ for the DBI global monopole can be easily obtained. For the $00$ and $11$ components, we have
\begin{equation}\label{Einstein 00}
\frac{1}{r^2}-\frac{1}{A}\left(\frac{1}{r^2}-\frac{1}{r}\frac{A^\prime}{A}\right)
= \epsilon^2 {T_0}^0,
\end{equation}
\begin{equation}\label{Einstein 11}
\frac{1}{r^2}-\frac{1}{A}\left(\frac{1}{r^2}+\frac{1}{r}\frac{B^\prime}{B}\right)
= \epsilon^2 {T_1}^1,
\end{equation}
where the corresponding components of the energy-momentum tensor of the DBI global monopole configuration reads
\begin{equation}\label{T0}
{T_0}^0 = \frac{1}{\lambda}\left(\frac{1}{\gamma}-1\right)+V_0 \left(f^2-1\right)^2,
\end{equation}
\begin{equation}\label{T1}
{T_1}^1 = -\frac{\gamma {f^\prime}^2}{A}+\frac{1}{\lambda}\left(\frac{1}{\gamma}-1\right)+V_0
\left(f^2-1\right)^2,
\end{equation}
with $\epsilon^2 = 8\pi G \sigma_0^2$, which is also a dimensionless constant. In principle, the metric functions $A(r)$, $B(r)$ and the DBI scalar field magnitude $f(r)$ can be obtained by solving the combined system of Eqs.(\ref{E-L Eq}), (\ref{Einstein 00}) and (\ref{Einstein 11}). However, solving these coupled equations analytically is a formidable problem, even though there exists a general solution to Einstein equations for the static spherically symmetric metric (\ref{line element}). For the  energy-momentum tensor ${T_\mu}^\nu$ shown in Eqs.(\ref{T0})-(\ref{T1}), it reads:
\begin{eqnarray}\label{A ex}
A(r)^{-1}&=&1-\frac{\epsilon^2}{r}{\int_0}^r\left[\frac{1}{\lambda}\left(\frac{1}{\gamma}-1\right)+V_0\left(f^2-1\right)^2\right]r^2dr,
\end{eqnarray}
\begin{eqnarray}
B(r)&=&\frac{1}{A(r)}\exp\left[\epsilon^2{\int_\infty}^r
\left(f'^2r\gamma\right)dr\right].
\end{eqnarray}

\section{DBI global monopole}

Let us first discuss the asymptotic behavior of the field configuration.
Using Eqs.(\ref{E-L Eq})-(\ref{T1}), we can find the asymptotic
expression for functions $f(r)$, $A(r)$, and $B(r)$. In the region of $r\gg1$,
\begin{eqnarray}\label{series expansion}
f(r)&=&1-\frac{1}{4V_0}\left(\frac{1}{r}\right)^2+\frac{8\lambda V_0+2\epsilon^2-3}{32{V_0}^2}\left(\frac{1}{r}\right)^4+O(r^{-5}),\\
A(r)&=&\frac{1}{1-\epsilon^2}+\frac{2G\sigma_0M_\infty}{(1-\epsilon^2)}\frac{1}{r}+\left[\frac{\epsilon^2(1+2\lambda V_0)}{4V_0(1-\epsilon^2)^2}+\frac{4G^2{\sigma_0}^2{M_\infty}^2}{(1-\epsilon^2)^3}\right]\left(\frac{1}{r}\right)^2+O(r^{-3}),\\
B(r)&=&(1-\epsilon^2)-2G\sigma_0M_\infty\frac{1}{r}-\frac{(1+2\lambda V_0)\epsilon^2}{4V_0}\left(\frac{1}{r}\right)^2+O(r^{-3}),
\end{eqnarray}
where the constant $M_\infty$ will be discussed below. Just as that of a canonical global monopole \cite{Harari}, the asymptotic expansion for $f(r)$ is  influenced much more heavily by $V_0$, rather than the choice of the parameter $\epsilon$ and $\lambda$.
On the other hand, in the region of $r\ll 1$, $f(r)$, $A(r)$, and $B(r)$ can be expressed asymptotically as
\begin{eqnarray}\label{series expansion-2}
f(r)&=&f_1r+\frac{f_1}{60\lambda \left(1+2\lambda f_1^2\right) \sqrt{1+3\lambda f_1^2}} \left\{
8\epsilon ^2-24\lambda V_0-8\epsilon ^2 \sqrt{1+3\lambda f_1^2}(1-\lambda V_0)\right.\nonumber\\
&+&\left.\lambda f_1^2 \left[43\epsilon ^2-144\lambda V_0-22\epsilon^2\sqrt{1+3\lambda f_1^2}(1-\lambda V_0) \right]+3\lambda^2 f_1^4 \left(19 \epsilon ^2-72\lambda V_0\right)\right\}r^3+O(r^{5}),\\
A(r)&=&1+\frac{\epsilon ^2}{3\lambda}\left(\sqrt{1+3\lambda f_1^2}-1+\lambda V_0\right)r^2+O(r^4),\\
B(r)&=&1+\frac{\epsilon ^2}{6\lambda} \left(2-2\lambda V_0- \frac{1}{\sqrt{1+3\lambda f_1^2}}- \sqrt{1+3\lambda f_1^2}\right)r^2+O(r^4),
\end{eqnarray}
where the undetermined coefficient $f_1$ is characterized as the mass of the monopole, which can be fixed in the following numerical calculation.

In analog with Schwarzschild vacuum solution, the metric coefficient
$A(r)$ and $B(r)$ outside the core can be formally written as
\begin{eqnarray}\label{A}
A(r)^{-1}&=&1-\epsilon^2-\frac{2G\sigma_0M_A(r)}{r},
\end{eqnarray}
\begin{eqnarray}\label{B}
B(r)&=&1-\epsilon^2-\frac{2G\sigma_0M_B(r)}{r}.
\end{eqnarray}

As we have mentioned above that $f(r)$ should
approach to unity in the region $r\gg1$, the mass functions $M_A(r)$ and $M_B(r)$ will
also approach to a definite value far away from the core. Thus, from Eqs.(\ref{series
expansion})-(\ref{B}), one can obtain their asymptotic expansions:
\begin{eqnarray}
M_A(r)&=&M_{\infty}+{\pi\sigma_0\left(2\lambda+\frac{1}{V_0}\right)}\frac{1}{r}
-\frac{\pi\sigma_0}{6}\left[\left(2\lambda+\frac{1}{V_0}\right)^2-\frac{\epsilon^2}{V_0^2}\right]\left(\frac{1}{r}\right)^3+O(r^{-4}),\\
M_B(r)&=&M_{\infty}+{\pi\sigma_0\left(2\lambda+\frac{1}{V_0}\right)}\frac{1}{r}
-\frac{\pi\sigma_0}{6}\left[\left(2\lambda+\frac{1}{V_0}\right)^2+\frac{\epsilon^2-3}{2V_0^2}\right]\left(\frac{1}{r}\right)^3+O(r^{-4}),
\end{eqnarray}
where $M_{\infty}$ is defined as:
\begin{eqnarray}
M_{\infty}\equiv \lim_{r\longrightarrow\infty} M_A(r) .
\end{eqnarray}

\begin{figure}
\begin{center}
\includegraphics[width=0.5\textwidth]{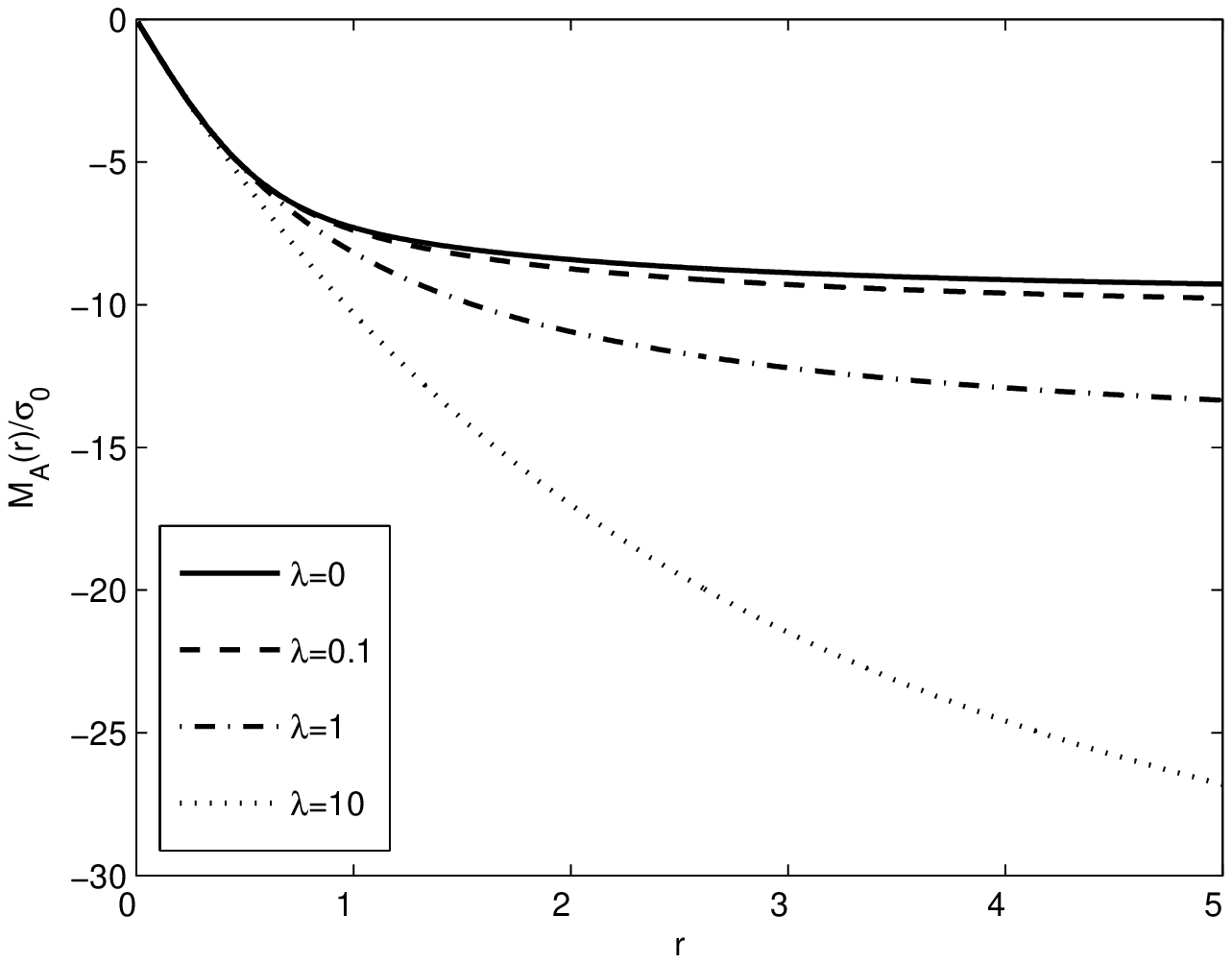}
\caption{The plot of $M_A(r)/\sigma_0$ as a function of $r$. Here we
choose $V_0= 1$  and $\epsilon = 0.001$. The three curves are plotted
in the condition that $\lambda = 0$ (which correspond to a canonical global monopole), $\lambda = 0.1$, $\lambda = 1$ and $\lambda= 10$,
respectively.}\label{fig:M}
\end{center}
\end{figure}

\begin{figure}
\begin{center}
\includegraphics[width=0.5\textwidth]{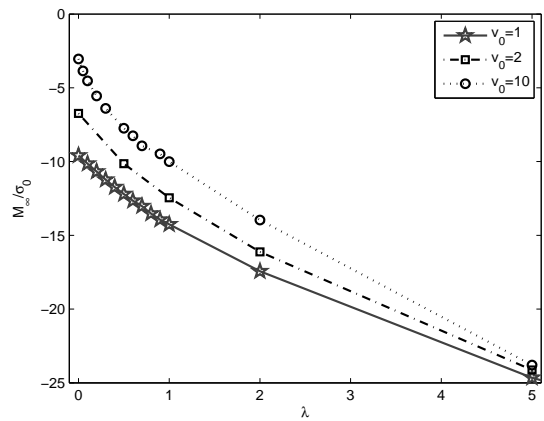}
\caption{The plot of $M_{\infty}/\sigma_0$ as a function of $\lambda$. Here we
choose $\epsilon = 0.001$ and the three curves are plotted
in the condition that $V_0 = 1,2$ and $10$,
respectively.}\label{fig:Minfty}
\end{center}
\end{figure}

\begin{figure}
\begin{center}
\includegraphics[width=0.5\textwidth]{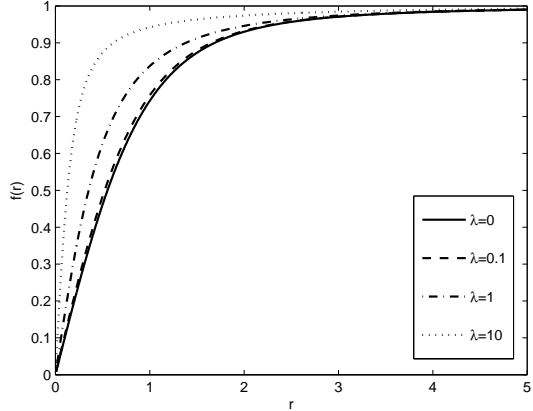}
\caption{The plot of $f(r)$ as a function of $r$. Here we
choose $V_0= 1$  and $\epsilon = 0.001$. The three curves are plotted
in the condition that $\lambda = 0$ (which correspond to a canonical global monopole), $\lambda = 0.1$, $\lambda = 1$ and $\lambda= 10$,
respectively.}\label{fig:f}
\end{center}
\end{figure}

The numerical results of $M_A(r)/\sigma_0$ are shown in
Fig.\ref{fig:M} by a shooting method for boundary value problems where
$V_0$ = 1 and $\epsilon$ = 0.001. It is easy to find from the figure
that the mass function of a DBI global monopole decreases from zero to a negative asymptotic
value $M_{\infty}$ when $r$ approach to infinity. Furthermore, the greater the value of parameter $\lambda$ takes,  the larger the absolute value of $M_{\infty}$ will become. This means that for the same conditions of potential height $V_0$ and the symmetry breaking scale $\sigma_0$, the effective mass of a DBI global monopole is less than that of a canonical global monopole.  The numerical values of $M_{\infty}/\sigma_0$ as a function of parameter $\lambda$ for different value of $V_0$ are plotted in Fig.\ref{fig:Minfty}. Furthermore, it is shown in Fig.\ref{fig:f} that $f(r)$ will trends to unity more quickly when the parameter $\lambda$ takes a larger value.

In fact, instead of solving the coupled equations(\ref{E-L
Eq})-(\ref{T1}), it is reasonable to  assume the following configuration
\begin{equation} \label{f assume}
f=\left\{ \begin{aligned}
         0\;\;\;\; \mathrm{if}\;\;\; r&<\delta,&\\
                 1\;\;\;\;\mathrm{if}\;\;\; r&>\delta,&
                          \end{aligned} \right.
                          \end{equation}
where $\delta$ denotes the core radius of the DBI global monopole. It should be noted that assumption (\ref{f assume}) is not an exact solution of Eq.(\ref{E-L Eq}). However, it will be shown later that,  taking such a simplified model for the monopole
configuration  is reasonable since it gives us a simple way  to discuss
the main features of the DBI global monopole. It is also worth pointing out that this
assumption  means that the monopole is modeled by a false
vacuum inside the core, and a true vacuum at the exterior. Thus, the
Einstein equation inside the core can be solved by de Sitter metric:
\begin{eqnarray}\label{inside}
ds^2 = \left(1-H^2r^2\right)dt^2-\frac{dr^2}{1-H^2r^2}-r^2d\Omega^2,
\end{eqnarray}
with $H^2 = {V_0\epsilon^2}/{3}$, while the exterior solution formally
reads
\begin{eqnarray}\label{outside}
ds^2 =
\left(1-\epsilon^2-\frac{2G\sigma_0M(r)}{r}\right)dt^2-\left({1-\epsilon^2-\frac{2G\sigma_0M(r)}{r}}\right)^{-1}{dr^2}-r^2d\Omega^2,
\end{eqnarray}
where $M(r)$ is the effective mass function of the DBI global monopole.
Comparing Eq.(\ref{A ex}) with Eq.(\ref{outside}) under the
assumption of Eq.(\ref{f assume}), for the region of $r>\delta$, one can obtain that
\begin{eqnarray}\label{M exact}
\frac{M(r)}{4\pi\sigma_0}=-r+\frac{1}{3}V_0\delta^3
-\frac{1}{3\lambda}\left(r^3-\delta^3\right)+\frac{1}{3\lambda}\left(2+r^2\right)^{3/2}
-\frac{1}{3\lambda}\left(2+\delta^2\right)^{3/2}.
\end{eqnarray}
Thus, using the continuity of the first derivatives of metric with
respect to radical proper distance $r$, we obtain the expression for
the core radius:
\begin{eqnarray}\label{del}
\delta=\sqrt{\frac{2}{V_0^2\lambda+2V_0}}.
\end{eqnarray}
For a canonical global monopole, which corresponds to the case that $\lambda=0$, the core size $\delta \propto V_0^{-1/2}$. Obviously, the existence of the $\lambda$-term shrinks the size of the core. Especially, for large value of $\lambda$ (i.e., $\lambda\gg V_0^{-1}$),  the size behaves asymptotically like $\delta \propto V_0^{-1}$.
Note that the size of the monopole core does not depend on the parameter $\epsilon$ and then the symmetric breaking scale $\sigma_0$.
If one defines the asymptotic value for $M(r)$, $M_\infty$, as
\begin{eqnarray}\label{M infinity}
M_\infty\equiv \lim_{r\longrightarrow\infty} M(r),
\end{eqnarray}
it is easy to find  that
\begin{eqnarray}\label{M infinity asymptotic}
M_\infty=-\frac{8\sqrt{2}\pi\left(1+\lambda V_0\right)}{3\sqrt{V_0(2+\lambda V_0)}}\sigma_0.
\end{eqnarray}
Clearly, since the value of parameters $V_0$, $\sigma_0$ and $\lambda$ are always greater than zero, the value of
$M_\infty$ is negative. In other words, just as that of a canonical global monopole, the effective mass of a DBI global monopole is actually negative. This conclusion is justified by the numerical results presented above. It is worth pointing out that when the condition $\lambda\ll V_0^{-1}$ is satisfied, $M_\infty\sim -V_0^{-1/2}\sigma_0$; however,  when $\lambda\gg V_0^{-1}$, $M_\infty\sim -\sqrt{\lambda}\sigma_0$, which does not depend apparently  on the value of $V_0$. Actually, to a certain extent, this property is already illustrated in Fig.\ref{fig:Minfty}, which shows that the values of $M_{\infty}$ for different values of $V_0$ are gradually close to each other as the parameter $\lambda$ grows larger.

Now let us turn to study the deflection of light around a DBI global monopole.
Since we are interested in the light moving outside the monopole
core, and the effective mass $M_A(r)$ and $M_B(r)$ approach to their
asymptotic value $M_\infty$, taking them as a constant $M_\infty$ is
a good approximation. We consider the geodesic equation in the
metric (\ref{line element}) with $A^{-1}(r) = B(r) = 1 - \epsilon^2
- \frac{2G\widetilde{M}}{r}$, where $\widetilde{M} = \sigma_0
M_\infty$. By introducing a dimensionless quality $\mu =
\frac{G\widetilde{M}}{r}$, it is easy to obtain a second order
differential equation with respect to $\mu$ :
\begin{eqnarray}
\frac{d^2\mu}{d\varphi^2}+(1-\epsilon^2)\mu = 3\mu^2.
\end{eqnarray}

Since $\frac{\widetilde{M}}{r}\ll1$, we can obtain the approximate
solution of $\mu$ :
\begin{eqnarray}
\mu =
G\widetilde{M}\left\{\frac{1}{l}\sin\left(\sqrt{1-\epsilon^2}\varphi\right)
+
\frac{G\widetilde{M}\left[1-\cos\left(\sqrt{1-\epsilon^2}\varphi\right)\right]^2}{l^2
(1-\epsilon^2)}\right\},
\end{eqnarray}
where $l$ is the closest approach to the core of the  monopole. Thus,
when light passes by the monopole, the deflect angle of it will
be
\begin{eqnarray}\label{angle}
\Delta\varphi = \frac{4G\widetilde{M}}{lc^2
(1-\epsilon^2)^{3/2}} \approx \frac{4G\widetilde{M}}{lc^2} +
\frac{6G\widetilde{M}\epsilon^2}{lc^2}
\end{eqnarray}

The last term in Eq.(\ref{angle}) is the modification when compared
with the deflect angle passing by an ordinary star. It should be
noted that since $\widetilde{M}$ is negative, this deflection
describes a force of repulsive property, not attractive. Moreover, since the effective mass of a DBI global monopole is more negative than that of a canonical one in the same condition of false vacuum and the symmetry breaking scale, this deflection is larger than that caused by a canonical global monopole.

\section{Conclusions}
Global monopole could be formed during the phase transition in the early
universe. In this paper,
we have generalized the field theory of global monopole to DBI scalar field and investigate the gravitational property of a DBI global monopole  in four-dimensional spherically symmetric spacetime.
Its main gravitational effect can be understood in terms of a deficit solid
angle, which reveals a relatively small potential of
repulsive nature. Compared with a canonical global monopole, for the same false vacuum and symmetry breaking scale, a DBI global monopole has a smaller core  and the absolute value of its effective mass which is always negative is larger. When the scale of warp factor $M$ becomes infinite or, in other words, the parameter $\lambda$ is close to zero, a DBI global monopole reduces to a canonical one and they are not distinct from each other in the circumstance. However, for large values of $\lambda$,  the effective mass of a DBI global monopole is not mainly determined by potential parameter $V_0$ but decided by $\lambda$, which is qualitatively different from that of a canonical or other type of non-canonical global monopole.

As a kind of $k$-monopole,  there certainly exist some common features between DBI monopole discussed here  and the $k$-monopole model considered in Ref.\cite{Jin}. For example, they both produce tiny gravitational field outside the
core in addition to a solid angular deficit, and both of the non-canonical kinetic terms in the two models can exert  significant influences on the effective mass of the monopole. However, the most obvious difference between them is that the effective mass of a $k$-monopole considered in Ref.\cite{Jin} can be positive, while that of a DBI monopole is always negative.

It is noteworthy that, throughout the paper, we only consider global monopole solution in a DBI scalar field theory in which the warp factor is assumed to be constant. As a matter of fact, in string/M theory, the effective DBI scalar field theory of D-brane usually contains a variable warp factor \cite{DBI}. For example, for a pure
$AdS_5$ throat, it is simply $\beta^4/\phi^4$ with $\beta$ a constant. It may be interesting to investigate global monopole solutions in these circumstances, which will be discussed in our preparing work.

\section*{Acknowledgments}
This work is supported in part by National Natural Science
Foundation of China under Grant  No. 10503002 and
Shanghai Commission of Science and Technology under Grant No.
06QA14039.

\end{document}